# Temperature dependence of contact resistance of Au–Ti–Pd$_2$Si–$n^+$-Si ohmic contacts

A.E. Belyaev[1], N.S. Boltovets[2], R.V. Konakova[1],

Ya.Ya. Kudryk[1], A.V. Sachenko[1], V.N. Sheremet[1]

[1]*V. Lashkaryov Institute of Semiconductor Physics, NAS Ukraine*

*41 Nauky Prospect, Kyiv 03028, Ukraine*

*Tel.: (380-44) 525-61-82; Fax: (380-44) 525-83-42; e-mail: konakova@isp.kiev.ua*

[2]*State Enterprise Research Institute "Orion", 8$^a$ Eugene Pottier St., Kyiv, 03057, Ukraine*

**Abstract.** We investigated temperature dependence of contact resistance of an Au–Ti–Pd$_2$Si ohmic contact to heavily doped $n^+$-Si. The contact resistance increases with temperature owing to conduction through the metal shunts. In this case, the limiting process is diffusion input of electrons to the metal shunts. The proposed mechanism of contact resistance formation seems to realize also in the case of wide-gap semiconductors with high concentration of surface states and dislocation density in the contact.

Paper received ; revised manuscript received ; accepted for publication .

In the last few years, some reports have appeared on anomalous temperature dependence of contact resistance $R_c$ (growth with temperature) of ohmic contacts to semiconductors with high dislocation density [1-3]. It should be noted that such behavior of $R_c(T)$ curves cannot be explained by the classical mechanisms of current transport (thermionic, thermal-field, tunnel) in the Schottky contacts. The authors of the above papers related the obtained $R_c(T)$ dependences to conduction through the metal shunts linked to dislocations.

In the present work, we describe growing with temperature dependences of contact resistance (measured in the 80–380 K temperature range) that take place in Au–Ti–Pd$_2$Si ohmic contacts to heavily doped silicon. To explain them, we also apply the mechanism of current transport through metal shunts. It is supposed that the limiting process is the mechanism of diffusion input of electrons to those metal shunts. The theoretical estimations showed that, in the case of heavily doped semiconductor material, such a limitation is possible only if accumulation band bending is realized in the semiconductor near the metal shunt ends. Such a situation seems rather realistic if one takes into account both the edge effect (leading to big increase of the electric field strength) and mirror image forces. In our case, not only a big decrease of barrier height near the shunt edge occurs but the band bending changes its sign as well.

We calculated contact resistance, with current limiting mechanism of diffusion input of electrons, by assuming that a rather high barrier is realized in the regions far from dislocations, so that it is possible to neglect current transport through the above regions. In that case, the contact resistance $R_c$ can be determined from the derivative of the current through all dislocations.

The density of the thermionic current $J_{nc}$ flowing through the contact at the site of dislocation outlet can be determined by solving the continuity equation for electrons. With allowance made for the diffusion limitation, the contact resistance related to a single dislocation, $R_{c0}$, is (in the case of a nondegenerate semiconductor)

$$R_{c0} = \frac{kT}{q} \frac{\left(1 + \frac{V_T}{4D_n} e^{y_{c0}} \alpha L_D\right)}{\frac{qV_T}{4} N_d e^{y_{c0}}}. \tag{1}$$

Here $k$ is the Boltzmann constant, $T$ temperature, $q$ electron charge, $V_T$ mean thermal velocity of electrons, $D_n$ electron diffusion coefficient, $y_{c0}$ dimensionless equilibrium band bending at the site of dislocation outlet, $\alpha$ numerical coefficient (of the order of unity), $L_D$ Debye shielding length, and $N_d$ donor concentration in the semiconductor.

If the semiconductor is degenerate and the current limitation due to the diffusion input takes place, one can also use Eq. (1). In that case, however, the following relation between the diffusion coefficient $D_n$ and electron mobility $\mu_n$ has to be used:

$$D_n = \frac{kT}{q} \mu_n \left( \frac{d(\ln n)}{dz} \right)^{-1}. \tag{2}$$

Here $z = E_f / kT$ is the dimensionless Fermi energy, and $n$ is the electron concentration in the semiconductor bulk:

$$n = \frac{2}{\sqrt{\pi}} N_c \left( \frac{T}{300} \right)^{1.5} \int_0^\infty \frac{\sqrt{x}}{1 + \exp(x-z)} dx, \tag{3}$$

where $N_c$ is the effective density of states in the conduction band at $T = 300$ K. The area from which the current flowing through a single dislocation is collected equals $\pi L_D^2$, where

$$L_D = \left( \frac{\varepsilon_0 \varepsilon_s kT}{2q^2 N_c} \right)^{0.5} \left( \Phi'_{1/2}(z) \right)^{-1/2} \tag{4}$$

is the Debye shielding length at an arbitrary degree of semiconductor degeneracy, $\varepsilon_0$ ($\varepsilon_s$) vacuum (semiconductor) permittivity, and

$$\Phi'_{1/2}(z) = \frac{2}{\sqrt{\pi}} \int_0^\infty \frac{\sqrt{x} \exp(x-z)}{(1+\exp(x-z))^2} dx. \tag{5}$$

The contact resistance for a contact of unit area determined by the mechanism of diffusion input is

$$R_{diff} = \frac{R_{c0}}{\pi L_D^2 N_{D1}}, \tag{6}$$

where $N_{D1}$ is the density of dislocations that take part in current transport. In general, $N_{D1}$ is not equal to the density of dislocations that take part in electron scattering, $N_{D2}$. The current transport is related to the dislocations that are normal to the contact plane, while scattering occurs at those dislocations that are oriented at an angle to the contact plane.

The quantity $\pi L_D^2 N_{D1} S$ ($S$ is the contact area) has a meaning of the total area from which the current flowing through all dislocations is collected. As a rule, the value of $\pi L_D^2 N_{D1}$ is much less than unity, even at maximal dislocation densities (about $10^{10}$–$10^{11}$ cm$^{-2}$), except the case of weakly doped semiconductors ($N_d \leq 10^{15}$ cm$^{-3}$).

The electron diffusion coefficient $D_n$ was determined with fitting procedure using Eqs. (2) and (3), while the temperature dependence of electron mobility $\mu_n$, when comparing the calculated and experimental dependences, was taken from [4] for the doping level of $2.8 \times 10^{19}$ cm$^{-3}$. The rated resistivity $\rho_{Si}$ of silicon samples used to make ohmic contacts was 0.002 Ω·cm. Taking into account its standard spread and the fact that predominant scattering at $T = 300$ K is that on the charged impurities (as a result, the dependence of $\rho_{Si}$ on the doping level is very weak), one can determine the limits for variation of donor concentration in the samples studied: $10^{19}$–$3 \times 10^{19}$ cm$^{-3}$. So we used both limiting values for donor concentration when plotting Fig. 1.

Shown in Fig. 1 are the experimental temperature dependence of contact resistivity for Au–Ti–Pd$_2$Si–$n^+$-Si ohmic contact c $\rho_{Si} = 0.002$ Ω·cm and $\rho_c(T)$ curves calculated for two donor concentrations: $10^{19}$ cm$^{-3}$ and $3 \times 10^{19}$ cm$^{-3}$. The contact metallization was made using layer-by-layer vacuum thermal sputtering of metals onto an $n^+$-silicon substrate heated up to 330 °C. The ohmic contact was formed by the Pd$_2$Si phase that appeared in a thin near-surface layer of $n^+$-Si in the course of palladium sputtering. Because of misfit of Pd$_2$Si and Si lattices and coefficients of thermal expansion, high density of structural defects (in particular, dislocations) is formed in the near-surface layer of $n^+$-Si [5-7]. According to [1-3], metal shunts (that short-circuit the space charge region) may form at the above defects. One should note very good agreement between the theory and experiment. It was obtained by using the general Eqs. (2)–(5) that take into account degeneracy of semiconductor.

The mechanism of contact resistance formation proposed in the present work for metal−semiconductor contacts has to realize, first of all, in the case of wide-gap semiconductors with high concentration of surface states in the contact. It looks rather paradoxical because current is transported through the regions accumulating electrons rather than the depletion regions. At the same time, the practically ideal agreement between the calculated and experimental temperature dependences of contact resistivity is indicative of the validity of the above mechanism. One should note that, in the model proposed in [1-3], the temperature dependence of contact resistance was strictly linear.

**Fig. 1.** The temperature dependences of contact resistivity $\rho_c$ for Au–Ti–Pd$_2$Si–$n^+$-Si ohmic contact (dots – experiment, curves – theory). Curve 1 (2) is plotted for $N_d = 10^{19}$ cm$^{-3}$ ($3\times10^{19}$ cm$^{-3}$). The values of parameters used in calculation: $V_T = 10^7$ cm/s; curve 1: $y_{c0} = 2.7$; $N_{D1} = 2.8\times10^9$ cm$^{-2}$; curve 2: $y_{c0} = 1.5$; $N_{D1} = 5\times10^9$ cm$^{-2}$.

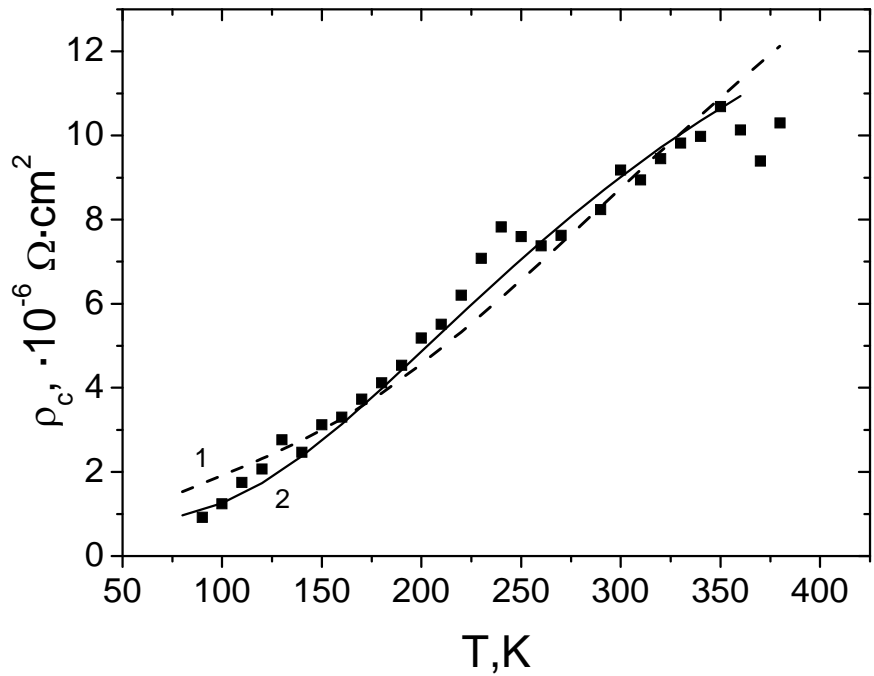

Fig. 1.